\documentclass[aps,prd,preprintnumbers,nofootinbib,floatfix]{revtex4}
\usepackage{xcolor}
\usepackage{graphicx}
\usepackage{wrapfig}
\usepackage{amsmath}
\usepackage{amsfonts}
\usepackage{amssymb}
\usepackage{multirow}
\usepackage{slashed}
\usepackage{physics}
\usepackage{float}
\usepackage{dcolumn}
\usepackage{color,soul}
\usepackage{comment}
\usepackage{multirow}
\usepackage[colorlinks=true, linkcolor=blue, citecolor=blue, urlcolor=blue]{hyperref}
\usepackage[textheight=9in, textwidth=6.5in, letterpaper]{geometry}

\newcommand\befs{\begin{figure*}}
\newcommand\eefs[1]{\label{fig:#1}\end{figure*}}
\newcommand\bef{\begin{figure}}
\newcommand\eef[1]{\vskip -0.125cm \label{fig:#1}\end{figure}}
\newcommand\beq{\begin{equation}}
\newcommand\eeq[1]{\label{#1}\end{equation}}
\newcommand\beqa{\begin{eqnarray}}
\newcommand\eeqa[1]{\label{#1}\end{eqnarray}}
\newcommand\bet{\begin{table}}
\newcommand\eet[1]{\label{tb:#1}\end{table}}
\newcommand\bets{\begin{table*}}
\newcommand\eets[1]{\label{tb:#1}\end{table*}}
\newcommand{\be}{\begin{equation}}
\newcommand{\ee}{\end{equation}}
\newcommand{\bea}{\begin{eqnarray}}
\newcommand{\eea}{\end{eqnarray}}

\newcommand\eqn[1]{Eq.\ (\ref{#1})}
\newcommand\scn[1]{Section \ref{sec:#1}}
\newcommand\apx[1]{Appendix \ref{sec:#1}}
\newcommand\tbn[1]{Table \ref{tb:#1}}

\begin{document}

\date{\today}

\title{Massless fermions in uniform flux background on $T^2\times R$: Vacuum quantum numbers from single-particle filled modes using lattice regulator}

\author{Nikhil\ \surname{Karthik}}
\email{nkarthik.work@gmail.com}
\affiliation{American Physical Society, Hauppauge, New York 11788}
\affiliation{Department of Physics, Florida International University, Miami, FL 33199}
\author{Rajamani\ \surname{Narayanan}}
\email{rajamani.narayanan@fiu.edu}
\author{Ray\ \surname{Romero}}
\email{rrome071@fiu.edu}
\affiliation{Department of Physics, Florida International University, Miami, FL 33199}

\begin{abstract} 
The quantum numbers of monopoles in $R^3$ in the presence of massless fermions  have been analyzed using 
a uniform flux background in $S^2\times R$ coupled to fermions. An analogous study in 
$T^2\times R$ is performed by studying the
discrete symmetries of the Dirac Hamiltonian in the presence of a static uniform field on $T^2$ with a total flux of $Q$ in the continuum. The degenerate ground states are classified based on their transformation properties under $\frac{\pi}{2}$ rotations of $T^2$ that leave the background field invariant. We find that the lattice analysis with overlap fermions exactly reproduces the transformation properties of the single particle zero modes in the continuum. Whereas the transformation properties of the single particle negative energy states 
can be studied in the continuum and the lattice, we are also able to study the transformation properties and the particle number (charge) of the many-body ground state on a finite lattice, and 
we show that the contributions from the fully filled single-particle states
cannot be ignored.

\end{abstract}
\maketitle
\newpage
\tableofcontents
\newpage

\section{Introduction}

Monopoles play a significant role in the modern analysis of quantum electrodynamics in three dimensions. When the gauge action suppresses monopoles, the resulting parity invariant quantum theory is scale invariant for all even number of flavors of massless two component fermions~\cite{Karthik:2015sgq,Karthik:2016ppr}. Computation of monopole scaling dimensions in an expansion in large number of flavors~\cite{Borokhov:2002ib,Pufu:2013vpa}; in an expansion around four dimensions~\cite{Chester:2015wao} and in conformal bootstrap~\cite{Chester:2016wrc,Li:2018lyb,Albayrak:2021xtd} suggests a critical number of flavors. This has been confirmed by numerical simulations~\cite{Karthik:2019mrr,Karthik:2024ffr}. A non-perturbative study of compact QED with massless fermions faces hurdles due to the proliferation of monopoles that are singular on the lattice. This leads to near-zero modes for Wilson fermions with negative masses, thereby causing a technical difficulty in numerical simulations with massless overlap fermions~\cite{Karthik:2019jds}. 

Numerical evaluation of the scaling dimensions of monopoles using a background field computation~\cite{Karthik:2019mrr,Karthik:2024ffr} does not address the question of the quantum numbers of the underlying monopole operators.
The determination of the scaling dimensions of monopole operators in the limit of a large number of flavors is a computation in a fermionic theory coupled to a static monopole background~\cite{Borokhov:2002ib}. For such an analysis, it is natural to consider radial quantization where $R^3 \to S^2\times R$.  The Euclidean time, $\tau$, is related to the radial coordinate by $r=e^{\tau}$. 
The Dirac operator in the presence of a spherically symmetric monopole was analyzed in~\cite{Kazama:1976fm}. The nonzero single-particle energy levels of the corresponding Hamiltonian are given by
\be
\epsilon = \pm \sqrt{p(p+|Q|)};\quad p\in \mathbf{N},
\ee
From this, the scaling dimensions of the monopole as a function of $Q$ can be read off from the corresponding Casimir energies of the ground states in the radial quantization~\cite{Borokhov:2002ib,Pufu:2013vpa}.

In addition to the scaling dimension, the monopole operators could carry non-trivial quantum numbers~\cite{Borokhov:2002ib,Pufu:2013vpa} as follows. All ground states are obtained by first filling the infinite number of negative energy states of the Hamiltonian. The vacuum state in the presence of monopoles is not unique due to the presence of zero modes. A single particle energy state labeled by an integer $p$ is also an eigenstate of the generalized angular momentum operator ${\bf K}$~\cite{Kazama:1976fm} with an eigenvalue $j=p+\frac{|Q|-1}{2}>0$; thus, in general, the single-particle states transform non-trivially under rotations.
Nevertheless, one assumes that the multiparticle vacuum state transforms trivially under ${\bf K}$, and one also assumes that the charge is zero under suitable normal ordering. With $Q=1$ the situation is reasonably clear~\cite{Borokhov:2002ib}. The zero mode has a total spin of zero. 
With $N$ flavors, one has $N$ zero modes, and
starting from the state obtained by filling all the negative energy states, we see that there are $\left({}^{N}_{k}\right)$ 
degenerate vacuum states obtained by filling the $k\in [0,N]$ zero energy modes. If we impose invariance under CP, the claim is that $N$ needs to be even ($N=2N_f$), and the states with $k=N_f$ are the ones that are invariant under CP. Since the spin of each of the zero modes is zero, we can conclude that all $\left({}^{2N_f}_{N_f}\right)$ will be rotationally invariant. The situation becomes more complicated when $|Q|>1$
and we have $2|Q|N_f$ zero modes. The spin of the zero mode is not zero, and only some of the $\left({}^{2|Q|N_f}_{|Q|N_f}\right)$ will have a total spin of zero. This has been analyzed in~\cite{Dyer:2013fja}.

This paper aims to understand the emergence of the quantum
numbers of the filled many-body vacuum states, now
in the presence of an
ultraviolet regulator.  This will help avoid the
assumptions about contributions from the infinite number of negative energy states.   Ideally, we would want to stay on $S^2\times R$ and use a lattice regularization but this is not simple to implement.  It is typical to implement the lattice regulator on 
the torus. Therefore, instead of addressing the issue in $S^2\times R$, 
we study the analogous problem of quantum numbers of vacua of massless fermions in the presence of a 
background static magnetic field $2\pi Q$ on $T^2$. This problem was also studied in~\cite{Song:2018ccm} where the focus was to study the effect of different types of bipartite lattices on the structure of the ground states. The study in this paper will be only on a square lattice, and we will use overlap fermions to realize a single two-component fermion. The spectrum of the Dirac Hamiltonian in this background has been analyzed in~\cite{Sachs:1991en}. The associated vacuum energy cannot be related to the monopole scaling dimensions, but a study of the quantum number of the degenerate states that form the vacuum is relevant. 

To be precise, we consider the Dirac Hamiltonian on a $L^2$ lattice with a background field that has a  uniform flux of $2\pi Q$. Spatial parity transformation is the same as charge conjugation for this particular gauge field background.
Overlap fermions for a single two-component fermion will have exactly $|Q|$ zero modes\footnote{Wilson fermions will also have $|Q|$ zero modes, but the Wilson mass parameter will have to be tuned to different values for each zero mode and all of these modes will coalesce only when the continuum limit is taken.}. Parity invariance is restored if we have a pair of flavors with $\left(H_o(Q), H_o(-Q)\right)$ as the associated Hamiltonians.
In this case, the two flavors put together will have $(2L^2-|Q|)$ negative eigenvalues and $2|Q|$ zero modes, which need to be suitably filled to create the multiparticle vacuum state. In this paper, we will address the transformation properties of such ground states under $\frac{\pi}{2}$ rotations. Our focus will be on the lowest and highest energy states that have a total of $L^2$ filled single-particle states, independent of $Q$. We will refer to this as the two half-filled sectors. The half-filling is motivated by the fact that the ground state of free fermions ($Q=0$) is unique (with anti-periodic boundary conditions on the fermions) and is made up of $L^2$ single-particle states, and we wish to maintain this condition for all $Q$ (imposing Gauss' law).

 The organization of the rest of the paper is as follows.  Our presentation in \scn{tormon} will be slightly unconventional -- we will set up the problem on the lattice and then take the continuum limit. We will discuss the action of the discrete rotation symmetry on the gauge fields. We will set up the Dirac Hamiltonian in \scn{dirham}. We will first analyze the problem in the continuum in \scn{contanal}. Much of the results will be taken out of~\cite{Sachs:1991en} but the presentation will be self-contained to help the reader. Particular attention will be paid to the transformation of the eigenstates (corresponding to both zero and non-zero eigenvalues) under $\frac{\pi}{2}$ rotations. Due to the infinite number of non-zero eigenvalues, we will not be able to properly address the issue of the transformation properties of the degenerate ground states. This will be resolved by repeating the study on the lattice in \scn{latanal}. We will show that the transformation properties of the zero modes exactly match the ones in the continuum even when $L$ is finite. This will help us isolate the role played by the transformation properties of the non-zero modes and show that a continuum statement can be made by studying the theory on finite $L$. In particular, we will address the flavor symmetry of the ground state sector at half-filling.

\section{Uniform field on a two-dimensional torus}\label{sec:tormon}

Let the size of the symmetric spatial torus be $\ell$ and we will refer to those spatial directions as $1,2$. The Euclidean time direction will be $3$.
To write down the gauge field in the continuum, we will start on a lattice where the periodic boundary conditions are imposed on the link variables. Let the periodic two-dimensional lattice be $L\times L$ and we label the points on the lattice by $(n_1,n_2)$  where the gauge links $U_i(n_1,n_2)$ satisfy
\be
U_i(n_1+k_1L,n_2+k_2L) = U_i(n_1,n_2);\qquad i=1,2
\ee
and $(k_1,k_2)$ being in the set of integers.
The link variables are naturally associated with the unitary parallel transporters, $T_k$, $k=1,2$ defined by
\be
[T_1\phi](n_1,n_2) = U_1(n_1,n_2) \phi(n_1+1,n_2); \qquad [T_2\phi](n_1,n_2) = U_2(n_1,n_2) \phi(n_1,n_2+1),
\ee
and $\phi(n_1,n_2)$ obey periodic boundary conditions, namely,
\be
\phi(n_1+L,n_2) = \phi(n_1,n_2+L) = \phi(n_1,n_2).
\ee
One choice for the gauge links that results in a uniform plaquette
\be
P(n_1,n_2) = U_1(n_1,n_2) U_2(n_1+1,n_2) U^*_1(n_1,n_2+1) U^*_2(n_1,n_2) = e^{i\frac{2\pi Q}{L^2}}\label{uniform}
\ee
is
\be
U_1(n_1,n_2) =\begin{cases} e^{-i\frac{2\pi Q n_2}{L}} & n_1=L-1 \cr 1 & 0\le n_1 < L-1.
\end{cases};\qquad 
U_2(n_1,n_2) = e^{i\frac{2\pi Qn_1}{L^2}}.\label{Qlat}
\ee
Explicit computation shows that $P(n_1,n_2)$ gives the expected value as long as $(n_1,n_2)\ne (L-1,L-1)$ and 
\be
P(L-1,L-1)  = e^{i\frac{2\pi Q}{L^2}} e^{-i2\pi Q}.
\ee
\eqn{uniform} is satisfied by all plaquettes if $Q$ needs to take on integer values  resulting in flux quantization on the torus.
The Polyakov loops are
\be
P_1(n_2) = \prod_{n_1=0}^{L-1} U_1(n_1,n_2) = e^{-i\frac{2\pi Q n_2}{L}};\qquad
P_2(n_1) = \prod_{n_2=0}^{L-1} U_2(n_1,n_2) = e^{i\frac{2\pi Q n_1}{L}}.
\ee

A rotation by $\frac{\pi}{2}$ of the lattice takes
\be
(n_1,n_2) \to (n'_1,n'_2) = ((L-n_2)\bmod L,n_1);
\ee
which results in $T'_k = R^t T_k R$, with the transformation matrix
\be
R_{n_1,n_2;n'_1,n'_2} = \delta_{n_1,n'_2}\left[ \delta_{n_2+n'_1,L}+\delta_{n_2+n'_1,0}\right];\qquad RR^t=\mathbf{I};\qquad R^4 =\mathbf{I}.\label{rotpi2}
\ee
The resulting link variables are
\be
U'_1(n'_1,n'_2) = e^{-i\frac{2\pi Qn'_2}{L^2}};\qquad
U'_2(n'_1,n'_2) =\begin{cases} e^{i\frac{2\pi Q n'_1}{L}} & n'_2=L-1 \cr 1 & 0\le n'_2 < L-1.
\end{cases},\label{rotlinks}
\ee
and we see that the plaquettes and Polyakov loops remain invariant. 
The above rotation is the same as the periodic gauge transformation
defined by
\be
G_{n_1,n_2;n'_1,n'_2} = e^{-i\frac{2\pi Q n_1 n_2}{L^2}}\delta_{n_1,n'_1}\delta_{n_2,n'_2};
\qquad G^\dagger G = \mathbf{I};\qquad
T'_k = G^\dagger T_k G.\label{gtrans}
\ee

The continuum limit is taken by introducing the lattice spacing $a$ and writing $x_i=an_i$ and $\ell = aL$. We see that 
$U_2(n_1,n_2)=e^{i a \frac{2\pi Q x_1}{\ell^2}} $ has a well-defined continuum limit but $U_1(n_1,n_2)$ is singular.
As a result, one modifies the periodic boundary conditions on the continuum field, $\phi$, such that
the gauge field in the continuum limit corresponding to the lattice fields in \eqn{Qlat} is written as
\be
A_1(x_1,x_2) =  0 ;\qquad 
A_2(x_1,x_2) = 
bx_1;\qquad b=\frac{2\pi Q}{\ell^2}
\label{hodgec}
\ee
along with 
\be
\phi(x_1+\ell,x_2) = e^{-i\frac{2\pi Q x_2}{\ell}}\phi(x_1,x_2);\qquad \phi(x_1,x_2+\ell) = \phi(x_1,x_2).\label{bcf}
\ee
The continuum gauge field associated with rotated lattice fields in \eqn{rotlinks} is written as
\be
x_1'=-x_2;\quad x_2'=x_1;\qquad A'_1(x'_1,x'_2)= -bx'_2,\quad A'_2=0
\ee
along with 
\be
\phi'(x'_1+\ell,x'_2) = \phi'(x'_1,x'_2);\qquad \phi'(x'_1,x'_2+\ell) = e^{i\frac{2\pi Q x'_1}{\ell}}\phi'(x'_1,x'_2).\label{bcfrot}
\ee
These are of course well known~\cite{Sachs:1991en}.

\section{Dirac Hamiltonian}\label{sec:dirham}

The Dirac Lagrangian is
\be
{\cal L} = \int d^2 x d\tau \ \bar\psi(x,\tau) \left[ \sigma_1 (\partial_1 +i A_1) + \sigma_2 (\partial_2 + iA_2) +\sigma_3 \partial_\tau\right] \psi(x,\tau),
\ee
with $\sigma_i$ being the standard Pauli matrices and  the total charge,
\be
J_0 = \int d^2 x\  \bar\psi(x,\tau) \sigma_3 \psi(x,\tau),
\ee
is time independent.
The time-independent multiparticle Hamiltonian is
\be
{\cal H} = 
\int d^2x \psi^\dagger(x,\tau) H \psi(x,\tau);\qquad
H = \begin{pmatrix} 0 & C \cr
C^\dagger  & 0\end{pmatrix};\qquad C=-(\partial_1+iA_1)+i(\partial_2+iA_2)\label{diracH}
\ee
and the total charge,
\be
J_0 = \int d^2 x\  \psi^\dagger (x,\tau) \psi(x,\tau),
\ee
is the number operator.

Lattice regularization will result in finite matrices for $H$ in \eqn{diracH}, and the integral over space will become a sum over lattice sites. This will enable us to perform a careful enumeration of the states along with their particle number (total charge). 
We will proceed to regularize $H$ using  overlap fermions~\cite{Narayanan:1993sk,Narayanan:1993ss,Narayanan:1994gw,Neuberger:1997fp} to ensure massless fermions without fine-tuning while avoiding doublers. The naive chiral Dirac operator is
\be
C = -\frac{1}{2}\left(T_1 -T_1^\dagger\right) +\frac{i}{2}\left(T_2-T_2^\dagger\right)\label{naive}
\ee
and the Wilson mass operator is
\be
B = 2 -m -  \frac{1}{2}\left(T_1 +T_1^\dagger\right) - \frac{1}{2}\left(T_2+T_2^\dagger\right);\qquad m \in (0,2).\label{wilson}
\ee
The overlap-Dirac   Hamiltonian is
\be
 H_o = \frac{ \sigma_3 + \epsilon(H_w)}{2};\qquad H_w = \begin{pmatrix} B & C \cr C^\dagger & - B \end{pmatrix}.\label{homat}
\ee
All eigenvectors of the unitary operator, $V=\sigma_3 \epsilon(H_w)$, satisfy the relation
\be
V \psi = e^{i\phi}\psi;\qquad V \sigma_3 \psi = e^{-i\phi}\sigma_3\psi.
\ee
The spectrum of $H_o$ is given by
\be
H_o \left[\frac{ \psi \pm e^{i\frac{\phi}{2}}\sigma_3 \psi}{\sqrt{2}}\right]=\pm \cos\frac{\phi}{2}
 \left[\frac{ \psi \pm  e^{i\frac{\phi}{2}}\sigma_3 \psi}{\sqrt{2}}\right].
\ee
If $Q\ne 0$, there will $|Q|$ zero modes of $H_o$ that are chiral, and their chiral pairs will have eigenvalues of $\pm 1$ depending on the sign of $Q$. 
\section{Discrete transformations}

Given the gauge field background in \eqn{Qlat}, the charge conjugated field is given by $U^c_\mu(n_1,n_2) = U_\mu(n_1,n_2)$ for $\mu=1,2$.
This takes $Q\to -Q$, and
let us denote the corresponding overlap-Dirac Hamiltonians by $H_o(Q)$ and $H_o(-Q)$ with $Q>0$. 

Given the gauge field background in \eqn{Qlat}, we define the field under spatial parity by
\be
n^p_1 = n_1;\qquad n^p_2={\rm mod}(L-n_2,L)
\ee
and this results 
\be
U^p_1(n^p_1,n^p_2) = U_1(n_1,L-n_2) = \begin{cases} e^{i\frac{2\pi Q n_2}{L}} & n_1=L-1 \cr 1 & 0\le n_1 < L-1.
\end{cases};\qquad 
U^p_2(n^p_1,n^p_2) = U^*_2(n_1,L-n_2)= e^{-i\frac{2\pi Qn_1}{L^2}}.
\ee
and we see that the resulting field is the same as the charge-conjugated field.

Our gauge field background is static, and we can therefore identify time reversal with particle $\leftrightarrow$ anti-particle.
Given the many-body Hamiltonian, ${\cal H} (Q) = a^\dagger H_o(Q) a$, a particle $\leftrightarrow$ anti-particle transformation
results in
\be
{\cal H}^\tau(Q) = a H_o(Q) a^\dagger = -a^\dagger H_o^t(Q) a
\ee
since $\Tr(H_o)=0$.
Noting that $B^t(Q)=B^*(-Q)$ and $C^t(Q) = C(-Q)$, we see that
\be
H_w^t(Q) = -\sigma_1 H_w(-Q) \sigma_1 \quad\Rightarrow\quad H_o^t(Q) = -\sigma_1 H_o(-Q) \sigma_1.
\ee
and therefore
\be
{\cal H}^\tau(Q) = a^\dagger \sigma_1 H_o(-Q) \sigma_1 a
\ee
implying this transformation is also related to charge conjugation. In essence, there is only one discrete transformation for this background field and we will refer to it as charge conjugation from now on.
Therefore, we will consider a pair of Hamiltonians, $\left(H_o(Q),H_o(-Q)\right)$ to ensure invariance under the discrete transformations and consider $N_f$ pairs of such flavors and consider ground states that are half-filled with the zero modes.
\section{Continuum analysis}\label{sec:contanal}

We start with the gauge field in \eqn{hodgec} which yields
\be
C= -\partial_1 -bx_1 +i\partial_2,
\ee
in \eqn{diracH}.
The Hamiltonian acts on two-component spinors, $\psi(x_1,x_2)$, that obey the non-trivial boundary conditions
\be
 \psi(x_1+l,x_2)=e^{-i\frac{2\pi Q x_2}{l}}\psi(x_1,x_2);\qquad   \psi(x_1,x_2+l) =
\psi(x_1,x_2),\label{fermionbc}
\ee
on the torus.
With
\be
\psi = \begin{pmatrix}\psi_+ \cr \psi_- \end{pmatrix}
\ee
the spectrum of the Hamiltonian is as follows for $Q\ne 0$:
\begin{itemize}
    \item {$Q>0$:}
The $Q$ normalized zero modes are given by
\be
 \psi^0_{j}(x_1,x_2) =   \begin{pmatrix} 0 \cr
\phi_{j,0}(x_1,x_2) 
\end{pmatrix};\qquad j\in [0,Q-1].
\ee
The paired normalized modes with eigenvalues $\pm \sqrt{\frac{4\pi Q }{\ell^2}n}$ and a degeneracy of $Q$ are
\be
 \psi^{\pm n}_{j}(x_1,x_2) = \frac{1}{\sqrt{2}} \begin{pmatrix} 
\phi_{j,n-1}(x_1,x_2)\cr
\mp \phi_{j,n}(x_1,x_2)
\end{pmatrix};\qquad j\in [0,Q-1].
\ee
    \item {$Q<0$:}
The $|Q|$ normalized zero modes are given by
\be
 \psi^0_{j}(x_1,x_2) =  \begin{pmatrix} 
\phi_{j,0}(x_1,x_2) \cr
0
\end{pmatrix};\qquad j\in [0,|Q|-1].
\ee
The paired normalized modes with eigenvalues $\pm \sqrt{\frac{4\pi |Q| }{\ell^2}n}$ and a degeneracy of $|Q|$ are
\be
 \psi^{\pm n}_{j}(x_1,x_2) = \frac{1}{\sqrt{2}} \begin{pmatrix} 
\phi_{j,n}(x_1,x_2)\cr
\mp \phi_{j,n-1}(x_1,x_2)
\end{pmatrix};\qquad j\in [0,|Q|-1].
\ee
\end{itemize}
The orthonormal functions $\phi_{j,n}(x_1,x_2)$ appearing in the above equations are given by
\bea
&& \phi_{j,n}(x_1,x_2) = \frac{(2\pi |Q|)^{\frac{1}{4}}}{\ell} \sum_{k=-\infty}^\infty e^{i\frac{2\pi (j+kQ)x_2}{\ell}} 
\phi_n\left(\sqrt{b}\left[ x_1 + \ell k +\frac{\ell j}{Q}\right]\right);\cr
&& \int_0^\ell dx_1 \int_0^\ell dx_2 \ \phi^*_{j',n'}(x_1,x_2)\phi_{j,n}(x_1,x_2) = \delta_{j,j'}\delta_{n,n'}.
\eea
and $\phi_n(y)$ are normalized eigenfunctions of the dimensionless one-dimensional Harmonic oscillator. All the details can be found in \apx{harosc}.

We proceed to first discuss the transformation of the single-particle wavefunction under rotations, keeping in mind that rotations changes the boundary conditions. Based on this, we will be able to 
discuss the transformation properties of the many-body vacuum.
The  Hamiltonian operator in \eqn{diracH} is
\be
H = -i\sigma_2\partial_1 +i\sigma_1(\partial_2+ibx_1)
\ee
and it acts on functions, $\psi(x_1,x_2)$, that obeys the boundary conditions in \eqn{bcf}.
Define the operator $G_H$ by
\be
G_H = \begin{pmatrix} G & 0 \cr 0 & G \end{pmatrix};\qquad 
G(x_1,x_2;y_1,y_2) = e^{-ibx_1x_2} \delta(x_1-y_1)\delta(x_2-y_2) ;\qquad G_H^\dagger G_H=\mathbf{I}
\label{gaugecont}
\ee
and $\phi=G_H^\dagger\psi$ obeys the boundary conditions in \eqn{bcfrot}. In addition,
\be
H' = G_H^\dagger HG_H = -i\sigma_2(\partial_1-ibx_2) +i\sigma_1\partial_2
\ee
acts on functions, $\phi(x_1,x_2)$.
We define the operator $R_H$ by
\be
R_H = \begin{pmatrix} iR & 0 \cr 0 & R \end{pmatrix};\qquad 
R(x_1,x_2;y_1,y_2) = \delta(x_1-y_2)\delta(x_2+y_1) \qquad R_H^\dagger R_H =\mathbf{I}.\label{rotcont}
\ee
and  note that
\be
H'=R_H^\dagger HR_H \quad\Rightarrow\quad H = (G_H R_H^\dagger )^\dagger H (G_H R_H^\dagger );\qquad
(G_H R_H^\dagger)^4 = \mathbf{I}.
\ee
Though rotation $R_H$ itself is not a symmetry of the 
Hamiltonian due to the nontrivial boundary condition, the transformation $G_H R_H^\dagger$ is a symmetry. Therefore, 
we can ask for the quantum numbers of the various single-particle states 
under $G_H R_H^\dagger$, which can take one of the values
$1,i,-1,-i$.

Using the result in \apx{pi2rot}, we have
\be
(R_H G_H^\dagger)\psi^{0,\pm n}_{j}(x_1,x_2)    
 = \begin{cases} i^n J_{jj'}(Q)\psi^{0,\pm n}_{j'}(x_1,x_2) & Q > 0 \cr
 i^{-n+1} J_{jj'}(Q)\psi^{0,\pm n}_{j'}(x_1,x_2) & Q < 0 
 \end{cases};\qquad J(-Q) = J^\dagger(Q).\label{controtJ}
\ee
There is an extra factor of $i$ between $Q>0$ and $Q<0$ in \eqn{controtJ}. 
\bet
\begin{tabular}{|c|c|c|c|c|}
\hline
$Q$ & $m_0$ & $m_1$ & $m_2$ & $m_3$ \cr
\hline
           1 &           1 &           0 &           0 &           0  \cr  
           2 &           1 &           0 &           1 &           0  \cr
           3 &           1 &           0 &           1 &           1  \cr
           4 &           2 &           0 &           1 &           1  \cr
           5 &           2 &           1 &           1 &           1  \cr
           6 &           2 &           1 &           2 &           1  \cr
           7 &           2 &           1 &           2 &           2  \cr
           8 &           3 &           1 &           2 &           2  \cr
           9 &           3 &           2 &           2 &           2  \cr
          10 &           3 &           2 &           3 &           2  \cr
          11 &           3 &           2 &           3 &           3  \cr
          12 &           4 &           2 &           3 &           3  \cr
\hline
\end{tabular}
\caption{Degeneracy table of the eigenvalues of $J$ for a fixed $Q$.}
\eet{fluxrep}
Let $m_j$; $j=0,1,2,3$ be the number of zero modes (after simultaneous diagonalization of $H$ and $G_H R_H^\dagger$) that transform as $1,i,-1,-i$ respectively for a given $Q$. \tbn{fluxrep} lists the eigenvalue degeneracies of the matrix $J$ for a fixed value of $Q>0$.  From \eqn{controtJ}, we see that the transformation matrix is $i J^\dagger(|Q|)$ for $Q < 0$ and therefore
\be
\{m_0,m_1,m_2,m_3\} \to \{m_1,m_0,m_3,m_2\}\quad{\rm under}\quad Q \to -Q.\label{fluxrepQtoQm}
\ee

Using above the properties of single-particle states,  we will now find the transformation properties of multiparticle ground states obtained by appropriately filling the single-particle modes. Let us assume $Q>0$ and consider ground states that are half-filled with the zero modes of $N_f$ pairs of $\left(H_o(Q),H_o(-Q)\right)$. Since we have an infinite number of negative and positive eigenvalues of $H_o(Q)$, we assume that the corresponding fully filled negative energy states of  $\left(H_o(Q),H_o(-Q)\right)$ are rotationally invariant.
Let $n_k$; $k=0,1,2,3$ be the number of zero modes transforming as $1,i,-1,-i$ respectively
that were filled to obtain a half-filled ground state.
Then
\be
\sum_{k=0}^3 n_k = QN_f;\qquad 0 \le n_0,n_1 \le (m_0+m_1)N_f;\qquad  0 \le n_2,n_3 \le (m_2+m_3)N_f.\label{gssol}
\ee
All solutions of $n_k;k=0,1,2,3$  can be obtained from the data in \tbn{fluxrep}.
For each solution to the above equation, the transformation under $\frac{\pi}{2}$ rotations is given by
\be
 (1)^{n_0} (i)^{n_1} (-1)^{n_2} (-i)^{n_3} \equiv e^{i\frac{\pi}{2}j};\qquad j=0,1,2,3.
 \label{groundtransC}
\ee
The number of such ground states will be
\be
d(\{n_k\},Q) = {(m_0+m_1)N_f \choose n_0}{(m_0+m_1)N_f \choose n_1}{(m_2+m_3)N_f \choose n_2}{(m_2+m_3)N_f \choose n_3}.\label{dimg}
\ee
The associated symmetry group is  $U(n_1)\times U(n_2)\times U(n_3)\times U(n_4)$ rotations of the zero modes and should be compared  with the maximally allowed $U(QN_f)$ rotations of the half-filled ground state.

\section{Lattice analysis}\label{sec:latanal}

We had an infinite number of negative energy states in the continuum, and we could only discuss the properties of filling the zero modes.  The situation under lattice regularization enables us to include the filled negative energy states in our discussion of the transformation properties. One should also note that, in the continuum, a highest energy many-body state obtained by filling all the positive modes and half of the zero modes is indistinguishable from the many-body ground state obtained by filling all the negative modes and the corresponding zero modes.
Therefore, we will discuss the transformation properties of the lowest and highest energy half filled state on the lattice
for $N_f$ pairs of $\left(H_o(Q),H_o(-Q)\right)$.

We have shown in \apx{latrot} that, like in the continuum, the symmetry of the lattice hamiltonian is a composition of rotation $R$ and a gauge transformation $G^\dagger$:
\be
R_H^\dagger H_o R_H = G^\dagger_H H_o G_H \quad\Rightarrow\quad H_o = (R_H G_H^\dagger)^\dagger H_o (R_H G_H^\dagger),
\ee
where $G_H$ and $R_H$ has the same structure as in \eqn{gaugecont} and \eqn{rotcont} and
$G$ and $R$ are given by \eqn{gtrans} and \eqn{rotpi2}.

Let $e^{i j^-_L \frac{\pi}{2}}$ be the phase by 
which a lattice multiparticle ground state transforms under $\pi/2$ rotation with $j^-_L=0,1,2,3$. Two terms now contribute to this phase --- net phase from 
the half-filled zero modes, and the net phase from all the 
fully filled negative modes. The latter contribution is obtained 
by the product of the eigenvalues of $G_H R_H^\dagger$ restricted 
to the filled negative modes of $H_o$, which is a determinant. Let $\det_-(Q)$ be this determinant of $G_H R_H^\dagger$ on the fully filled negative energy states.

Similarly, let $e^{i j^+_L \frac{\pi}{2}}$ be the phase by 
which a highest energy state transforms under $\pi/2$ rotation with $j^+_L=0,1,2,3$. Contribution to this phase now comes from 
the same half-filled zero modes, and the net phase from all the 
fully filled positive modes given 
by the product of the eigenvalues of $G_H R_H^\dagger$ restricted 
to the filled positive modes. Let $\det_+(Q)$ be this determinant of $G_H R_H^\dagger$ on the fully filled positive energy states.
Using the formula for $\det (G_HR_H^\dagger)$ in \eqn{detfull}, we can relate $\det_+(Q)$ to $\det_-(Q)$ via the relation
\be
\det_+(Q) \det_+(-Q) \det_-(Q) \det_-(-Q) i^{|Q|} = (-1)^L,
\ee
where the factor of $i^{|Q|}$ is the contribution from the zero modes~\footnote{ The factor, $i^{|Q|}$, comes from \tbn{fluxrep} and \eqn{fluxrepQtoQm}:\ $
\left[1^{m_0} i^{m_1} (-1)^{m_2} (-i)^{m_3}\right]\left[ 1^{m_1} i^{m_0} (-1)^{m_3} (-i)^{m_2}\right] = i^{m_0+m_1+m_2+m_3}=i^{|Q|}.$}.

\bet
\begin{tabular}{|c|c|c|c|c|c|c|c|c|}
\hline
\multirow{2}{*}{$Q$}  & \multicolumn{3}{c|}{$\det_-(Q)$} & \multicolumn{3}{c|}{$\det_-(-Q)$} & \multirow{2}{*}{$\det_-(Q)\det_-(-Q)$}\\ \cline{2-7}
& $L=8,12$ & $L=10,14$ & $L=9,11,13,15$ &  $L=8,12$ & $L=10,14$ & $L=9,11,13,15$ & \cr
\hline
1 & i & -i & 1& -i & i & 1 & 1 \cr
2 & 1 & -1 & -1 & -1 & 1 & 1 & -1 \cr
3 & -1& 1 & -i &  i & -i & 1 & -i \cr
4 & -i & i & -i & 1 & -1 & 1 & -i \cr
5 & -i & i & -1 & -i & i & 1 & -1 \cr
6 & -1 & 1 & 1 & -1 & 1 & 1 & 1 \cr
7 & 1 & -1 & i & i & -i & 1 & i \cr
8 & i & -i & i & 1 & -1 & 1 & i \cr
9 & i & -i & 1 & -i & i & 1 & 1 \cr
10 & 1 & -1 & -1 & -1 & 1& 1 & -1 \cr
11 & -1 & 1 & -i & i & -i & 1 & -i \cr
12 & -i & i & -i & 1 & -1 & 1& -i \cr
\hline
\end{tabular}
\caption{Table of representations of the fully filled negative energy states on the lattice for positive and negative $Q$.}
\eet{fluxrepdet}

Given a value of $L\in [8,15]$ and $\pm Q \in [1,12]$, we numerically evaluated the eigenvalues and eigenvectors of $H_o$. The number of zero modes are equal to $|Q|$ and their chiralities are $-\frac{Q}{|Q|}$ like in the continuum. These are paired with $|Q|$ modes of opposite chiralities with eigenvalues $\frac{Q}{|Q|}$. The number of paired positive and negative eigenvalues away from $0,\pm 1$ are $L^2-|Q|$ and these are not chiral.
The transformation properties of the zero modes (eigenvalues of $G_HR_H^\dagger$) turn out to be independent of $L$ for $L>8$ and exactly matches the continuum values obtained from \eqn{controtJ}. The results for $\det_-(Q)$ are tabulated in \tbn{fluxrepdet}.  
We note from \tbn{fluxrepdet} that $\det_-(Q) \det_-(-Q)$ is independent of $L$ even though each of the factors depend on $L$ and we empirically find
\be
\det_-(Q) \det_-(-Q) = e^{i\frac{\pi(|Q|-1)(6-|Q|)}{4}} \quad\Rightarrow\quad \det_+(Q) \det_+(-Q) = (-1)^L e^{i\frac{\pi(|Q|+1)(|Q|-10)}{4}}.
\ee

With this information in place, one can list the transformation properties under $\frac{\pi}{2}$ rotation of all multiparticle ground states and the highest energy states of $N_f$ pairs $(H_o(Q), H_o(-Q))$. The fully filled negative and positive energy states transform trivially if $N_f$ is a multiple for $4$.
For $N_f$ not a multiple of $4$, the transformation properties of the fully filled negative and positive energy states of $(H_o(Q),H_o(-Q))$ become relevant. Since the transformation properties of the zero modes under $\frac{\pi}{2}$ rotations are found to be the same on the lattice and continuum, the phase from the zero modes is still $e^{i j\frac{\pi}{2}}$ with integer $j$ as defined in \eqn{groundtransC} for the continuum case. The phase from the fully filled modes is 
$\left[ \det_\pm(Q) \det_\pm(-Q)\right]^{N_f}$. Putting everything together, multiparticle ground states and the highest energy states of the lattice regularized Hamiltonian transform under $\pi/2$ rotation by
\be
e^{i j^\pm_L\frac{\pi}{2}} = e^{i\frac{\pi}{2}j} \left[ \det_\pm(Q) \det_\pm(-Q)\right]^{N_f};\quad{\rm where}\quad e^{i\frac{\pi}{2}j}=  \left[1^{n_0} i^{n_1} (-1)^{n_2} (-i)^{n_3}\right],\label{groundtransL}
\ee
where the choice of $\pm$ corresponds to the lowest and highest states respectively. As before, $n_k$; $k=0,1,2,3$ are the number of zero modes transforming as $1,i,-1,-i$ respectively that were filled to obtain the half-filled ground state. One should note 
two points:
\begin{enumerate}
\item The determinant factors are not identity (except for $N_f=4$), and thus the fully filled states contribute to the transformation properties of the multiparticle vacua. Given a solution to \eqn{gssol}, the effect of the fully filled positive or negative energy states is simply to map the associated $j$ in \eqn{groundtransC} to a $j^\pm_L(j,Q,L)$ on the lattice. 
\item $\det_+(Q)\det_+(-Q) \ne \det_-(Q)\det_-(-Q)$, and thus on regularization, the ground-state and the highest excited states start behaving differently.
\end{enumerate}

\section{Conclusions}

A classification of the monopole quantum numbers in three-dimensional QED is usually done by considering the fermion spectrum in the background of a spherical monopole. This is a local operator and the transformation properties of the zero modes can be obtained by a study of the generalized angular momentum operator. One typically used radial quantization, $r=e^\tau$, where $\tau$ is the Euclidean time and the $r=0$ is mapped to $\tau=-\infty$. The Hamiltonian is defined on $S^2$ and the monopole has a constant flux on each $S^2$ in $\tau=(-\infty,\infty)$. The quantum numbers as extracted from $S^2\times R$ computation could be obtained by computation of two point functions on the lattice with periodic boundary conditions. However, it is difficult to regulate the $S^2$ computations directly on the lattice. Therefore, we considered an analogous problem of uniform flux on $T^2$ and studied the spectrum of the fermion Hamiltonian in order to gain understanding of how the vacuum quantum numbers arise when lattice regulator is used. 
Unlike $S^2$ where the monopole is invariant under a continuous rotation, we only have discrete $\frac{\pi}{2}$ rotations on $T^2$. Such a study is of relevance from the viewpoint of Dirac spin liquids~\cite{Song:2018ccm,Song:2018ial} where the transformation of the monopole operators are considered on lattices with different discrete symmetries. In addition to the transformation properties of the zero modes of the lattice operator under $\frac{\pi}{2}$ rotations, we also need to look at the transformation properties of all eigenmodes since we look at the full transformation property of the multi-particle ground states and this includes filling all negative energy states of the Hamiltonian. Considering $2N_f$ flavors ($N_f$ parity invariant pairs), we found that the transformation properties of the half-filled many body state on the lattice is not the same as the one in the continuum. The set of continuum states that transform as $e^{i\frac{\pi}{2}j}$ in the continuum along with the associated symmetry group gets non-trivially mapped to a $j^-_L(j,Q)$ when we consider the lowest energy state and to a $j^+_L(j,Q,L)$ when we consider the highest energy state.
To make sure that our lattice observations are not due to lattice artifacts,
we used exactly massless overlap fermion which preserves the flavor
symmetry even at finite lattice spacings, and studied the 
vacuum quantum numbers at multiple values of lattice spacings at fixed background fluxes.
Even though we reduced the continuous
symmetry group on $S^2$ to the discrete group rotation on $T^2$,
we have shown that we cannot ignore the transformation property of
the fully filled negative energy state. Furthermore, we were able
to take the continuum limit by including the transformation of the
full vacuum state and not just the contribution of the zero modes
as is done in a continuum analysis on $S^2\times R$. Given the observed
significance of negative energy states on the torus, it would be
important to regulate the theory on $S^2\times R$ and study the
continuum limit of the transformation properties.

\acknowledgments
 R.N. acknowledges partial support by the NSF under grant number
PHY-1913010 and PHY-2310479.

\appendix
\section{Spectrum of the Dirac Hamiltonian in the background of a uniform flux on $T^2$}\label{sec:harosc}
The problem of the spectrum of the Dirac Hamiltonian in the background of a uniform flux on $T^2$ has been solved in~\cite{Sachs:1991en} and our aim is to study the transformation of the eigenvectors under $\frac{\pi}{2}$ rotations. To make the presentation self-contained,
we write down the details of the eigenvalue problem here.
 The eigenvalue problem for the Hamiltonian is
\be
(-\partial_1 -bx_1 +i\partial_2)\psi_-(x_1,x_2;E) = E\psi_+(x_1,x_2;E);\qquad (\partial_1 -bx_1 +i\partial_2)\psi_+(x_1,x_2;E) = E\psi_-(x_1,x_2;E).
\ee
We expand both $\psi_\pm$ in a momentum basis in the $x_2$ direction and write
\be
\psi_\pm(x_1,x_2;E) = \sum_{k=-\infty}^\infty c^k_\pm e^{ipx_2} \phi_\pm(y;E);\qquad  p=\frac{2\pi k}{\ell};\qquad y=\frac{1}{\sqrt{|b|}}\left( bx_1 + p\right).
\ee
Then
\be
\psi_\pm(x_1+\ell,x_2;E) =  e^{-\frac{2\pi i Qx_2}{\ell}}\sum_{k=-\infty}^\infty c^{k-Q}_\pm e^{ipx_2} \phi_\pm\left(y;E\right).
\ee
To match with the boundary condition in the $x_2$ direction, we conclude
\be
 c_\pm^k = c_\pm^{k-Q},
 \ee
 and the degeneracy of each energy eigenstate is $Q$ fold.
With $E=\sqrt{2|b|}\alpha $,
the differential equations reduce to
\bea
\sum_{k=-\infty}^\infty e^{ipx_2} \frac{1}{\sqrt{2}}\left[ \frac{Q}{|Q|} \frac{d}{dy} + y\right]c^k_-  \phi_-(y;E) &=& -\alpha\sum_{k=-\infty}^\infty e^{ipx_2} c_+^k\phi_+(y;E);\cr
\sum_{k=-\infty}^\infty e^{ipx_2} \frac{1}{\sqrt{2}}\left[ -\frac{Q}{|Q|}\frac{d}{dy} + y\right]c^k_+  \phi_+(y;E) &=& -\alpha\sum_{k=-\infty}^\infty e^{ipx_2} c_-^k\phi_-(y;E).
\eea
The connection to a one-dimensional Harmonic oscillator is evident, and
we use the standard creation and annihilation operators,
\be
a=\frac{1}{\sqrt{2}}\left[\frac{d}{dy}+y\right];\qquad a^\dagger=\frac{1}{\sqrt{2}}\left[-\frac{d}{dy}+y\right],
\ee
and label the states by real valued functions, $\phi_n(y)$; $n=0,1,\cdots$, such that
\be
a\phi_n =\sqrt{n}\phi_{n-1};\quad a^\dagger\phi_n = \sqrt{n+1}\phi_{n+1};\qquad
\int_{-\infty}^\infty dy\ \phi_n(y) \phi_{n'}(y) = \delta_{n,n'};\quad \phi_n(-y) = (-1)^n \phi_n(y).\label{horel}
\ee
Let us define
\bea
&& \phi_{j,n}(x_1,x_2) = \frac{(2\pi |Q|)^{\frac{1}{4}}}{\ell} \sum_{k=-\infty}^\infty e^{i\frac{2\pi (j+kQ)x_2}{\ell}} 
\phi_n\left(\frac{b}{\sqrt{|b|}}\left[ x_1 + \ell k +\frac{\ell j}{Q}\right]\right);\quad j\in [0,|Q|-1]\cr
&& \Rightarrow \begin{cases} 
a \phi_{j,n} = \sqrt{n}\phi_{j,n-1};\qquad a^\dagger \phi_{j,n} = \sqrt{n+1}\phi_{j,n+1};\cr
\int_0^\ell dx_1 \int_0^\ell dx_2 \ \phi^*_{j',n'}(x_1,x_2)\phi_{j,n}(x_1,x_2) = \delta_{j,j'}\delta_{n,n'}
\end{cases}.\label{phijn}
\eea
We will separately analyze the cases for $Q>0$ and $Q<0$.
\begin{itemize}
    \item {$Q>0$:}
In this case,
\be
c_-^k a \phi_-(y;E) = -\alpha c_+^k \phi_+(y;E);\qquad c_+^k a^\dagger \phi_+(y;E) = -\alpha c_-^k \phi_-(y;E).
\ee
We have $Q$ zero modes and this corresponds to $\phi_-=\phi_{j,0}$ and $\phi_+=0$ and the normalized zero modes are given by
\be
 \psi_{j,0}(x_1,x_2) =   \begin{pmatrix} 0 \cr
\phi_{j,0}(x_1,x_2)
\end{pmatrix};\qquad j\in [0,Q-1].
\ee
The paired normalized modes with eigenvalues $\pm \sqrt{\frac{4\pi Q }{\ell^2}n}$ each have a degeneracy of $Q$ and are
\be
 \psi^\pm_{j,n}(x_1,x_2) = \frac{1}{\sqrt{2}} \begin{pmatrix} 
\phi_{j,n-1}(x_1,x_2)\cr
\mp \phi_{j,n}(x_1,x_2)
\end{pmatrix};\qquad j\in [0,Q-1].
\ee
    \item {$Q<0$:}
In this case,
\be
c_-^k a^\dagger \phi_-(y;E) = -\alpha c_+^k \phi_+(y;E);\qquad c_+^k a \phi_+(y;E) = -\alpha c_-^k \phi_-(y;E).
\ee
We have $|Q|$ zero modes and this corresponds to $\phi_+=\phi_{j,0}$ and $\phi_-=0$ and  
the normalized zero modes are given by
\be
 \psi_{j,0}(x_1,x_2) =  \begin{pmatrix} 
\phi_{j,0}(x_1,x_2) \cr
0
\end{pmatrix};\qquad j\in [0,|Q|-1].
\ee
The paired normalized modes with eigenvalues $\pm \sqrt{\frac{4\pi |Q| }{\ell^2}n}$ each have a degeneracy of $Q$ and are
\be
 \psi^\pm_{j,n}(x_1,x_2) = \frac{1}{\sqrt{2}} \begin{pmatrix} 
\phi_{j,n}(x_1,x_2)\cr
\mp \phi_{j,n-1}(x_1,x_2)
\end{pmatrix};\qquad j\in [0,|Q|-1].
\ee
\end{itemize}

\subsection{Transformation of continuum eigenvectors under $\frac{\pi}{2}$ rotations}\label{sec:pi2rot}

Let us start with $\phi_{j,n}$ in \eqn{phijn} and note that 
$G^\dagger \phi_{j,n}$ obeys the boundary conditions in \eqn{bcfrot}. Therefore, we can write
\be
\sum_{k=-\infty}^\infty e^{i\frac{2\pi Q x_1 x_2}{\ell^2}}e^{i\frac{2\pi (kQ+j)x_2}{\ell}}
 \phi_n\left(\frac{b}{\sqrt{|b|}}\left[ x_1 + \ell k +\frac{\ell j}{Q}\right] \right)= \sum_{k'=-\infty}^\infty e^{i\frac{2\pi k'x_1}{\ell}} f_{j,n}(x_2,k').
 \ee
Then
\bea
f_{j,n}(x_1,k') &=& \frac{1}{\ell} \int_0^\ell \ dx_1 \sum_{k=-\infty}^\infty e^{i\frac{2\pi}{\ell^2} \left[  Q x_1 x_2+ (kQ+j)x_2\ell -  k'x_1\ell \right]}
 \phi_n\left(\frac{b}{\sqrt{|b|}}\left[ x_1 + \ell k +\frac{\ell j}{Q}\right] \right)\cr
 &=& \frac{1}{\ell} \int_0^\ell \ dx_1\sum_{k=-\infty}^\infty e^{i\frac{2\pi}{\ell^2} \left[  (Q x_2-lk')\left( x_1+ k\ell +\frac{\ell j}{Q}\right) +\ell^2 k'k +\ell^2 \frac{k'j}{Q}\right]}
 \phi_n\left(\frac{b}{\sqrt{|b|}}\left[ x_1 + \ell k +\frac{\ell j}{Q}\right] \right)\cr
 &=& \frac{e^{i\frac{2\pi k'j}{Q}}}{\ell} \int_{-\infty}^\infty \ dx_1 \ e^{ib x_1\left(x_2 -\frac{lk'}{Q}\right)}\phi_n\left(\frac{b}{\sqrt{|b|}} x_1 \right) \cr
 &=& \frac{e^{i\frac{2\pi k'j}{Q}}}{\sqrt{|b|} \ell} \int_{-\infty}^\infty \ dy \ e^{i\frac{b}{|b|}xy}\phi_n(y);\qquad x= \frac{b}{\sqrt{|b|}}\left(x_2-\frac{lk'}{Q}\right);\qquad y =\frac{b}{\sqrt{|b|}} x_1
\eea
Let us define
\be
\chi_n(x) = i^{-n} \int_{-\infty}^\infty \ dy \ e^{ixy}\phi_n(y)\quad\Rightarrow\quad \chi^*_n(x) = \chi_n(x);\quad \chi_n(-x) = (-1)^n\chi_n(x).
\ee
Then
\bea
\frac{d\chi_n(x)}{dx} &=& i^{-n+1} \int_{-\infty}^\infty \ dy \ e^{ixy} y\phi_n(y)\cr
&=& \begin{cases} i^{-n+1} \int_{-\infty}^\infty \ dy \ e^{ixy}\left( \sqrt{2n} \phi_{n-1}(y) -\frac{d\phi_n}{dy}\right) \cr
i^{-n+1} \int_{-\infty}^\infty \ dy \ e^{ixy}\left( \sqrt{2(n+1)} \phi_{n+1}(y) +\frac{d\phi_n}{dy}\right)
\end{cases}\cr
&=& \begin{cases} i^{-n+1} \int_{-\infty}^\infty \ dy \ e^{ixy}\left( \sqrt{2n} \phi_{n-1}(y) +ix \phi_n \right) \cr
i^{-n+1} \int_{-\infty}^\infty \ dy \ e^{ixy}\left( \sqrt{2(n+1)} \phi_{n+1}(y) -ix \phi_n\right)
\end{cases}\cr
&=& \begin{cases} \sqrt{2n}\chi_{n-1}(x) -x \chi_n(x)\cr
-\sqrt{2(n+1)}\chi_{n+1}(x) +x\chi_n(x)\end{cases},
\eea
where we have used the recursion relations for $\phi_n(y)$ from \eqn{horel}
and we see that $\chi_n(x)$ satisfy the same recursion relations, namely,
\be
\frac{d\chi_n(x)}{dx} + x\chi_n(x) = \sqrt{2n}\chi_{n-1}(x);\qquad -\frac{d\chi_n(x)}{dx} + x \chi_n(x) = \sqrt{2(n+1)}\chi_{n+1}(x).
\ee
To fix the constant relating $\chi_n(x)$ and $\phi_n(x)$, we note that
\be
\chi_0(x) = \frac{1}{\sqrt{\pi}} \int_{-\infty}^\infty e^{ixy-\frac{y^2}{2}} dy =
\frac{e^{-\frac{x^2}{2}}}{\sqrt{\pi}} \int_{-\infty}^\infty e^{-\frac{1}{2}(y-ix)^2} dy = \sqrt{2\pi} \phi_0(x) \quad\Rightarrow\quad \chi_n(x) = \sqrt{2\pi} \phi_n(x).
\ee
Therefore, 
\be
(G^\dagger\phi_{j,n})(x_1,x_2) = i^{\frac{b}{|b|}n} \sum_{j'=0}^{Q-1} \frac{e^{-i\frac{2\pi jj'}{Q}}}{\sqrt{|Q|}} \phi_{j',n}(x_2,-x_1) = i^{\frac{b}{|b|}n }\sum_{j'=0}^{Q-1} \frac{e^{-i\frac{2\pi jj'}{Q}}}{\sqrt{|Q|}} (R^t\phi_{j',n})(x_1,x_2)
\ee
resulting in 
\be
(RG^\dagger\phi_{j,n})(x_1,x_2) = i^{\frac{b}{|b|}n} \sum_{j'=0}^{Q-1} J_{j,j'}(Q)\phi_{j',n}(x_1,x_2);\qquad
J_{j,j'}(Q)=\frac{e^{-i\frac{2\pi jj'}{Q}}}{\sqrt{|Q|}} .
\ee

We note that
\be
J J^\dagger = \mathbf{I}
\ee
showing that it is a unitary matrix.
Next, we note that
\be
[J^2]_{jj'} = \frac{1}{|Q|} \sum_{j"} e^{-i\frac{2\pi(j+j')j"}{Q}}=\delta(j+j',0) + \delta(j+j',Q)
\quad\Rightarrow\quad J^4=\mathbb{I}.
\ee

\section{Gauge transformations and $\frac{\pi}{2}$ rotations on the lattice}\label{sec:latrot}

We start with the rotation matrix in \eqn{rotpi2} and consider
\bea
\left(R^t [ T_1 \phi]\right)(m_1,m_2) &=& R^t_{m_1,m_2;k_1,k_2} [T_1\phi](k_1,k_2)
= \begin{cases} [T_1\phi](m_2,0) & m_1=0 \cr [T_1\phi](m_2,L-m_1) & 0 < m_1 \le L-1.
\end{cases}\cr
&=& \begin{cases} U_1(m_2,0)\phi(m_2+1,0) & m_1=0 \cr U_1(m_2,L-m_1)\phi(m_2+1,L-m_1) & 0 < m_1 \le L-1\end{cases}
\cr
&=& \begin{cases} U_1(m_2,0)(R^t\phi)(0,m_2+1) & m_1=0 \cr U_1(m_2,L-m_1)(R^t\phi)(m_1,m_2+1) & 0 < m_1 \le L-1.
\end{cases}
\eea
\bea
\left(R^t [ T_2 \phi]\right)(m_1,m_2) &=& R^t_{m_1,m_2;k_1,k_2} [T_2\phi](k_1,k_2)
= \begin{cases} [T_2\phi](m_2,0) & m_1=0 \cr [T_2\phi](m_2,L-m_1) & 0 < m_1 \le L-1.
\end{cases}\cr
&=& \begin{cases} U_2(m_2,0)\phi(m_2,1) & m_1=0 \cr U_2(m_2,L-m_1)\phi(m_2,L-m_1+1) & 0 < m_1 \le L-1.
\end{cases}\cr
&=& \begin{cases} U_2(m_2,0)(R^t\phi)(L-1,m_2) & m_1=0 \cr U_2(m_2,L-m_1)(R^t\phi)(m_1-1,m_2) & 0 < m_1 \le L-1.
\end{cases}.
\eea
If we now define
\be
R^tT_1R = T'_2;\qquad R^t T_2 R = {T'_1}^\dagger;\qquad \chi=R^t\phi
\ee
then
\be
U'_1(n_1,n_2) =  U_2^*(n_2,L-1-n_1) \qquad
U'_2(n_1,n_2) = \begin{cases} U_1(n_2,0) & n_1=0;\cr U_1(n_2,L-n_1) & 0 < n_1\le L-1\end{cases}.
\ee
The above transformation matches \eqn{rotlinks}.
Referring to \eqn{naive} and \eqn{wilson} we have
\be
R^t C R = iC';\qquad R^t B R = B'
\ee
where $C'$ and $B'$ are in terms of $T'_i$; $i=1,2$.
Then
\be
R_H = \begin{pmatrix} iR & 0 \cr 0 & R \end{pmatrix}\quad
\Rightarrow\quad R^\dagger_H H_w R_H = H'_w \quad
\Rightarrow\quad R^\dagger_H \epsilon(H_w) R_H = \epsilon(H'_w).
\ee

Writing the gauge transformation using \eqn{gtrans} as
\be
G_H = \begin{pmatrix} G & 0 \cr 0 & G \end{pmatrix}
\quad\Rightarrow\quad
H'_w = G^\dagger_H H_w G_H \quad
\Rightarrow\quad G^\dagger_H \epsilon(H_w) G_H = \epsilon(H'_w).
\ee
In addition,
\be
\sigma_3 = R_H^\dagger \sigma_3 R_H = G^\dagger_H \sigma_3 G_H.
\ee
Therefore,
\be
H'_o = R_H^\dagger H_o R_H = G_H^\dagger H_o G_H.
\ee

Next we write,
\be
(G_H R^\dagger_H)_{n_1,n_2;m_1,m_2} = e^{-i\frac{2\pi Q n_1 n_2}{L^2}} \delta_{n_1,m_2}\left[ \delta_{n_2+m_1,0}+\delta_{n_2+m_1,L}\right]
\ee
Then
\be
(G_H R^\dagger_H)^2_{n_1,n_2;m_1,m_2}=e^{-i\frac{2\pi Q n_1 (n_2+m_2)}{L^2}}\left[ \delta_{n_1+m_1,0}+\delta_{n_1+m_1,L}\right]
\left[ \delta_{n_2+m_2,0}+\delta_{n_2+m_2,L}\right],
\ee
and
\bea
(G_H R^\dagger_H)^4_{n_1,n_2;n'_1,n'_2}&=&e^{-i\frac{2\pi Q n_1 (n_2+m_2)+m_1(m_2+n'_2)}{L^2}}\left[ \delta_{n_1+m_1,0}+\delta_{n_1+m_1,L}\right]
\left[ \delta_{n_2+m_2,0}+\delta_{n_2+m_2,L}\right]\cr
&&\left[ \delta_{m_1+n'_1,0}+\delta_{m_1+n'_1,L}\right]
\left[ \delta_{m_2+n'_2,0}+\delta_{m_2+n'_2,L}\right].
\eea
If $n_1=0$ ( we have $m_1=n'_1=0$) or $n_2=0$ (we have $m_2=n'_2=0$) and the phase prefactor is unity.
If $n_1\ne 0$ and $n_2\ne 0$, then $n_1+m_1=m_1+n'_1=n_2+m_2=m_2+n'_2=L$ and the prefactor is $e^{-i2\pi Q}=1$.
Therefore, we conclude that
\be
(G_H R_H^\dagger)^4=1;\qquad (G_H R_H^\dagger)^\dagger (G_H R_H^\dagger)=1,
\ee
and the eigenvalues of $G_HR_H^\dagger$ are $\pm 1,\pm i$. It can be shown that 
\be
\det (G_HR_H^\dagger) = \begin{cases}
       \text{$-i$,} &\quad\text{for L odd}, \\
       \text{$\left(-1\right)^{Q}$,} &\quad\text{for L even}
     \end{cases}.\label{detfull}
\ee
Note that if we combine a pair $\left(H_o(Q),H_o(-Q)\right)$, the result is independent of $Q$ but depends on $L$. One may absorb this dependence by redefining $R_H$ with an $L$ dependent sign. 
\bibliography{biblio}
\end{document}